\relax
\documentstyle [art12]{article}
\begin {document}
\bibliographystyle {plain}

\title{\bf Superconducting Instability in a One-Dimensional
Lattice of Berry Phase
Molecules}
\author {D. Shelton and A. M. Tsvelik}
\maketitle
\begin {verse}
$Department~ of~ Physics,~ University~ of~ Oxford,~ 1~ Keble~ Road,$
\\
$Oxford,~OX1~ 3NP,~ UK$\\
\end{verse}
\begin{abstract}
\par
We study the one-dimensional model of conduction electrons interacting
with local vibronic modes. States in the  conduction band are
two-fold
degenerate both in orbital index and spin. It is shown that
such one-dimensional system has a strong tendency for
superconductivity. The superconducting pairing originates from
coupling of two charge density waves;
the scaling dimension of the superconducting order
parameter is found  to be close to 1/4.
\end{abstract}

\sloppy
\par

 The discovery of superconductivity with
fairly high transition temperatures  in the alkali-intercalated fullerides
A$_{n}$C$_{60}$ (A = K, Rb and Cs) has revived an  interest to the Jahn-Teller
phenomenon. It is widely believed that the origin of superconducting pairing
in the C$_{60}$ compounds is due to the interaction of electrons with
the vibronic modes of C$_{60}$ molecules. A nontrivial aspect  of this
interaction is  an orbital degeneracy of the local
electronic states. In the strong coupling limit when one neglects the
intermolecular electron
hopping, each C$_{60}$ molecule is described as a system of
three local vibronic modes
interacting  with a three-fold degenerate electronic orbital. This
interaction is non-diagonal thus leading  to violations of the
adiabatic approximation (the problem of Berry phase) and, as a
consequence, to a strong renormalization
of the electronic wave functions (the Dynamical Jahn-Teller effect; see
Ref. 1 for a review).

 It is interesting to know whether a local degeneracy can survive in
the presence of an intermolecular hopping and if yes, how it will
affect the low-energy dynamics of the electronic subsystem. One
obvious choice would be  a  structure where C$_{60}$ molecules are
situated on chains.
In this case the one-dimensional electronic
band is split into a single degenerate band created by the $s, p_z$ orbitals
and a doubly degenerate band created by $p_x, p_y$ orbitals. The
measurements performed for RbC$_{60}$ give many indications of
a quasi-one-dimensional behaviour$^2$. The direct
band structure calculations done for A$_{n}$C$_{60}$-compounds with
A = K, Rb show, however,  that the three-fold degeneracy is lifted with the
characteristic energy of the order of 0.5$eV$ and the resulting
electronic bands have an essentially three dimensional character$^3$.
However, even if the results of the band theory calculations are correct,
the C$_{60}$-based materials are not very far from being
one-dimensional. The authors of Ref. 4
remark that the interchain hopping is very
sensitive to the distance between the chains of C$_{60}$ and therefore
by alloying the alkali with
``spacer molecules'' such as NH$_3$ one can achieve a higher degree
of one-dimensionality.

 It is well known that in the standard one-dimensional
attractive Hubbard model
the superconductivity competes with the charge density wave
instability. At small coupling constant the scaling dimension of the
superconducting order parameter is close to 1. This is clearly seen
for the the simpliest exactly solvable model of fermions with a
point-like interaction (see, for example, Ref. 4, where the relevant
correlation functions for this model are listed).
As we shall demonstrate in this paper, the superconducting pairing
occurs in our model between electrons from different orbitals. Thus
the presence of orbital
symmetry is crucial in this respect. We shall also demonstrate that
the scaling dimension of the order parameter is 1/4, that is fairly small.
A similar result was obtained for
a one-dimensional lattice of Berry phase molecules
in
the strong coupling limit by Manini {\it et al}.$^{5}$ and Santoro
{\it et al.}$^{6}$.

 In this paper we study the model introduced in Ref. 5; it describes
an one-dimensional
chain of Jahn-Teller molecules with electrons belonging
to two degenerate local orbitals hopping  along the chain and
interacting with local optical (vibronic) modes of the molecules.
The Hamiltonian
is given by

\begin{eqnarray}
H = \sum_r \sum_{\alpha = \pm 1}\sum_{a = 1,2}
\{ - \frac{1}{2}t[c^+_{\alpha, a}(r + 1) c_{\alpha, a}(r)
+ c^+_{\alpha, a}(r) c_{\alpha, a}(r + 1)]
\nonumber\\
+ J \left(c^+_{r,\alpha, a}\hat{\sigma^x}_{ab}c_{r,\alpha, b}X_r +
c^+_{r,\alpha a}\hat{\sigma^y}_{ab}c_{r,\alpha b}Y_r\right)] \nonumber\\
= \sum_r\left( - \frac{1}{2M}\frac{\partial^2}{\partial {X_r}^2} -
\frac{1}{2M}\frac{\partial^2}{\partial {Y_r}^2} +  \frac{kX_r^2}{2} +
\frac{kY_r^2}{2}\right) \label{eq:model}
\end{eqnarray}
Here $c^+_{\alpha,a}(r), \: c_{\alpha,a}(r)$ are creation and
annihilation of electrons on the cite $r$ and $n_a =
\sum_{\alpha}c^+_{\alpha,a}c_{\alpha,a}$.
The Greek indices correspond to a spin projection
and the English ones denote two
degenerate orbitals. The electrons interact with local vibronic modes
$X_r, \: Y_r$. As we have mentioned above, a
single C$_{60}$ molecule has a higher degeneracy,
but we assume  that such high
degeneracy does not  survive in a crystal.

 Instead of  the adiabatic approximation which is a standard
tool in the theory of the Jahn-Teller effect we shall use another
approach. Namely, we neglect
all retardation effects related to the kinetic energy of vibrons and
integrate over the vibronic modes. As in the standard BCS theory, this
will give us a high energy cut-off of the order of the vibronic
frequency $\omega_0 = (k/M)^{1/2}$. As the result we obtain the
following effective interaction between the fermions:
\begin{equation}
V_{\mbox{int}} = - \frac{1}{2}g\sum_r[(c^+_r\sigma^xc_r)^2 +
(c^+_r\sigma^yc_r)^2] = -
g\sum_rc^+_{1,\alpha}c_{1,\beta}c^+_{2,\beta}c_{1,\alpha}
\end{equation}
where $g = J^2/k$.

 Assuming  that $g/t << 1$ we  linearize
the electronic spectrum and then apply the standard bosonization procedure.
At the first step
we represent the fermionic operators as sums of right and
left moving components
\begin{equation}
c_{r} = R(x)e^{- \mbox{i}k_Fx} + L(x)e^{\mbox{i}k_Fx}
\end{equation}
where $x = ra$ ($a$ is the lattice constant). Substituting this into Eq.(2) and
keeping only non-oscillatory terms
we get
\begin{eqnarray}
V_{\mbox{int}} = V_1 + V_2 + V_3\\
V_1 =  g[R^+_{1,\:\alpha}R_{1,\:\beta}L^+_{2,\:\beta}L_{2,\:\alpha} +
R^+_{2,\:\alpha}R_{2,\:\beta}L^+_{1,\:\beta}L_{1,\:\alpha}] \\
V_2 = - g[L^+_{1,\:\alpha}R^+_{2,\:\beta}R_{1,\:\beta}L_{2,\:\alpha} +
L^+_{2,\:\alpha}R^+_{1,\:\beta}R_{2,\:\beta}L_{1,\:\alpha}]\\
V_3 =  g[R^+_{1,\:\alpha}R_{1,\:\beta}R^+_{2,\:\beta}R_{2,\:\alpha} + (R
\rightarrow L)] \label{eq:inter}
\end{eqnarray}

 We proceed further by bosonizing the fermionic operators:
\begin{eqnarray}
R_{a,\:\sigma}, \: R^+_{a,\:\sigma} = \frac{1}{\sqrt{2\pi a_0}}\exp(
\pm\mbox{i}\varphi_{a,\:\sigma}),\nonumber\\
L_{a,\:\sigma} , \: L^+_{a,\:\sigma}= \frac{1}{\sqrt{2\pi a_0}}\exp(
\mp\mbox{i}\bar\varphi_{- a,\:\sigma}) \label{eq:bos}\\
R^+_{a,\:\sigma}R_{a,\:\sigma} =
\frac{1}{\pi}\partial_z\Phi^a_{\sigma}, \: \:
L^+_{a,\:\sigma}L_{a,\:\sigma} = \frac{1}{\pi}\partial_{\bar
z}\Phi^a_{\sigma}\nonumber\\
\Phi(z, \bar z) = \varphi(z) + \bar\varphi(\bar z), \: \:\Theta(z, \bar
z) = \varphi(z) - \bar\varphi(\bar z)
\end{eqnarray}
where $a_0 = \Lambda^{-1}$ is the cut-off and $a = 1,2$ is the orbital
index,
$\sigma = \pm 1$ is the spin index and $z = \tau +
\mbox{i}x, \: \bar z = \tau -
\mbox{i}x$. In order to simplify the Hamiltonian we introduce bosonic
variables for charge (c) and spin (s) degrees of freedom on each
orbital by
\begin{eqnarray}
\varphi_{a,\:\sigma} = \frac{1}{\sqrt 2}(\varphi_{a,\mbox{c}} +
\sigma\varphi_{a,\mbox{s}})
\end{eqnarray}
and then the combinations corresponding to total (+) and relative (-)
charge and spin density fluctuations:
\begin{eqnarray}
\varphi_{\pm,\mbox{c,s}} = \frac{1}{\sqrt 2}\left(\varphi^{1}_{\mbox{c,s}}
\pm \varphi^{2}_{\mbox{c,s}}\right)
\end{eqnarray}
The bosonized Hamiltonian has the following form:
\begin{eqnarray}
H = H_0 + V_1 + V_2 + V_3\nonumber\\
H_0 =
\frac{1}{2}\sum_{\pm}\left[v_cK^{\pm}_c(\partial_x\Theta^{\pm}_c)^2 +
\frac{v_c}{K^{\pm}_c}(\partial_x\Phi^{\pm}_c)^2 +
v_sK^{\pm}_s(\partial_x\Theta^{\pm}_s)^2 +
\frac{v_s}{K^{\pm}_s}(\partial_x\Phi^{\pm}_s)^2\right]\label{ho}\\
V_1 = \frac{g}{(\pi a_0)^2}\cos\Phi_s^+\cos\Theta_s^-\nonumber\\
V_2 = \frac{g}{(\pi a_0)^2}[\cos\Phi_c^-\cos\Phi_s^- +
\cos\Phi_c^-\cos\Theta_s^-]\nonumber\\
V_3 = \frac{g}{(\pi a_0)^2}\cos\Phi_s^-\cos\Theta_s^- \label{intr}
\end{eqnarray}
The bosonization of $V_1$ and $V_3$ in (\ref{eq:inter}) also generates
the following terms:
\begin{equation}
V_4 = \frac{g}{2\pi^2}\left[(\partial_x\Phi^{+}_c)^2 -
(\partial_x\Phi^{-}_c)^2 +
(\partial_x\Phi^{+}_s)^2 - (\partial_x\Phi^{-}_s)^2\right]
\end{equation}
which renormalize the velocities $v_{c,s}$ and the exponents
$K^{\pm}_{c,s}$ from their bare values to
\[
\tilde K^{\pm}_{c,s} = (1 \pm g/\pi^2v_{c,s})^{-1/2}K^{\pm}_{c,s}
\]
They can therefore be absorbed into the bare Hamiltonian
(\ref{ho}). The resulting Hamiltonian is a particular case of the
Hamiltonian considered by Khveshchenko and Rice in relation to the
problem of two coupled Hubbard chains$^7$. Therefore we can
use the  renormalization group equations derived in this work. We
simplify these equations assuming that deviations of $K$'s from 1 are
small ($K^{\pm}_{c,s} = 1 \mp x^{\pm}_{c}$) and expanding to the
lowest nontrivial order in $x$:
\begin{eqnarray}
\dot g_1 = (2 - K_s^+ - \frac{1}{K_s^-})g_1 \approx (x_s^+ +
x_s^-)g_1\nonumber\\
\dot g_2 = (2 - K_c^- - K_s^-)g_2 \approx (x_c^- + x_s^-)g_2\nonumber\\
\dot g_3 = (2 - K_c^- - \frac{1}{K_s^-})g_3 \approx (x_c^- - x_s^-)g_3
\end{eqnarray}
where $g_1$ refers to the term  $V_1$ and  $g_2$ and $g_3$ to the
first and the second term of $V_2$ respectively. The interaction $V_3$
is irrelevant. The dots stand for differentiation with respect to $\xi
= \ln(\Lambda/\max(\omega, q))$. The equations for the exponents are
\begin{eqnarray}
\dot K_c^+ = 0, \: \dot K_c^- = - \frac{1}{2}(K_c^-)^2(g_2^2 +
g_3^2)\nonumber\\
\dot K_s^+ = - \frac{1}{2}(K_s^+)^2g_1^2, \: \: \dot K_s^- = -
\frac{1}{2}[(K_s^-)^2g_2^2 - g_1^2 - g_3^2]
\end{eqnarray}
We note that the first equation is exact in all orders since the
Hamiltonian is quadratic in the
$\Phi_c^+$-field.  To the first order in $x_{c,s}^{\pm}$ we have,
defining $x_1 = (x_s^+ + x_s^-), \: x_2 = (x_c^- + x_s^-), \: x_3 =
(x_c^- - x_s^-)$ :
\begin{eqnarray}
\dot g_i = x_ig_i\\
2\dot x_1 = 2g_1^2 - g_2^2 + g_3^2, \: 2\dot x_2 = 2g_2^2 -
g_1^2\nonumber\\
2\dot x_3 = g_1^2 + 2g_3^2 \label{sys}
\end{eqnarray}
In order to determine the character of the renormalization group flow
it is convenient to rewrite this system using the variables $y_i =
\ln(g_i^2)$ such that $\dot y_i = 2x_i$. Then Eqs.(\ref{sys}) become
\begin{eqnarray}
\ddot y_1 = 2\mbox{e}^{2y_1} - \mbox{e}^{2y_2} +
\mbox{e}^{2y_3}\nonumber\\
\ddot y_2 = 2\mbox{e}^{2y_2} - \mbox{e}^{2y_1}\nonumber\\
\ddot y_3 = \mbox{e}^{2y_1} + 2\mbox{e}^{2y_3}
\end{eqnarray}
The right hand sides of all these equations never vanish
simultaneously. Therefore
we conclude that coupling constants grow. Thus
we can be sure that the system scales to strong coupling.

 From the bahaviour of $x_s^+, \: x_{c,s}^-$ we conclude that in the
strong coupling regime gaps open in
the spectrum of $\Phi^+_s, \: \Phi^-_{c,s}$ and the corresponding
complex exponents acquire non-zero expectation values. From the form
of the interaction (\ref{intr}) it is evident  that non-zero expectation
values are acquired by exponents of the
fields $\Phi^-_{c}, \: \Phi^+_{s}, \: \Theta^-_s$. The field
$\Phi^+_c$ remains a free field. This situation is very advantageous
for the superconductivity. Indeed, the superconducting order parameter
\begin{eqnarray}
\Delta = R_{1, \:+}L_{2, - } \sim \exp\left[\frac{\mbox{i}}{2}(
\Phi^-_{s} + \Theta^-_{s} + \Phi^+_{s} + \Theta^+_{c})\right]
\nonumber\\
\sim  m^2:\exp\left(- \frac{\mbox{i}}{2}\Theta^+_{c}\right):
\end{eqnarray}
and the asymptotics of the two-point correlation function at distances
greater than $m^{-1}$ is  given by
\begin{eqnarray}
\langle \Delta_a(1)\Delta^+_b(2)\rangle \sim m^{3/2}[\tau_{12}^2 +
x_{12}^2]^{-1/4K^+_c}\nonumber\\
\frac{1}{K^+_c} = (1 + \frac{g}{\pi^2v_c})^{1/2}
\end{eqnarray}

 At small $g/v_c$ this function behaves almost
like the correlation function of hard-core
bosons (the transverse Pauli matrices in the XY-model). Since the
latter model has the same central charge $c = 1$, we conclude that
the gapless excitations in the
present model are well described by the model of hard-core bosons. Thus our
results are in complete agreement with the results of Santoro et al.$^6$
obtained within a different approach. Such a small scaling dimension
increases the  probability of having  a large transition temperature in
a quasi-three-dimensional system built of weakly interacting
chains. If $t$ is  the interchain Joshephson coupling,
then  the dimensional analysis gives for
the temperature of superconducting transition the estimate
\[
T_{\mbox{c}} \sim m(t/m)^{\frac{1}{2 - 2d}}
\]
In the conventional case $d \approx 1$ and the transition temperature
is exponentially small in $m/t$, but for the present model  $d = 1/4K^+_c$ and
at small $g/v_c$ we have $T_{\mbox{c}} \sim t^{2/3}$.

  The authors are
grateful to M. C. M. O'Brien and E. Tosatti for the interest to the
work and G. Santoro for sending his preprint.

\end{document}